%% file: main.tex
\documentclass[conference]{IEEEtran}

\usepackage{cite}

\usepackage[acronym]{glossaries} 
\input{acronyms}

\usepackage{csquotes}

\usepackage{pbalance}

\usepackage{listings}
\lstdefinestyle{listingstyle}{
    basicstyle=\ttfamily\footnotesize,        
    breaklines=true,       
    frame=lines,
    numbers=none
}
\usepackage[table]{xcolor}
\definecolor{light-gray}{gray}{0.95}

\usepackage{makecell}

\usepackage{multirow}

\usepackage{mdframed}
\usepackage[breakable, many]{tcolorbox} 

\newtcolorbox{answerbox}{
    breakable,
    colback = sub, 
    boxrule = 0pt, 
    leftrule = 0pt,
}

\usepackage{url}

\usepackage{graphicx}

\usepackage{orcidlink}

\hypersetup{
colorlinks=true,
linkcolor=black,
citecolor=black,     
urlcolor=black
}

\begin{document}

\input{text.tex}

\bibliographystyle{IEEEtran}
\bibliography{bibliography}

\end{document}

%% file: acronyms.tex
\newacronym{ai}{AI}{artificial intelligence}
\newacronym{altai}{ALTAI}{Assessment List for Trustworthy Artificial Intelligence}
\newacronym{api}{API}{application programming interface}
\newacronym{bom}{BoM}{bill of materials}
\newacronym{sdoc}{SDoC}{supplier's declaration of conformity}
\newacronym{slr}{SLR}{systematic literature review}

%% file: text.tex
\title{Model Cards Revisited: Bridging the Gap Between Theory and Practice for Ethical AI Requirements}

\author{
\IEEEauthorblockN{Tim Puhlfürß \orcidlink{0000-0001-8421-8071}}
\IEEEauthorblockA{\textit{University of Hamburg}\\  
Hamburg, Germany\\
tim.puhlfuerss@uni-hamburg.de}
\and
\IEEEauthorblockN{Julia Butzke \orcidlink{0009-0008-4184-8987}}
\IEEEauthorblockA{\textit{University of Hamburg}\\  
Hamburg, Germany}
\and
\IEEEauthorblockN{Walid Maalej \orcidlink{0000-0002-6899-4393}}
\IEEEauthorblockA{\textit{University of Hamburg}\\  
Hamburg, Germany\\
walid.maalej@uni-hamburg.de}
}


\maketitle

\begin{abstract}
Model cards are the primary documentation framework for developers of \gls{ai} models to communicate critical information to their users.
Those users are often developers themselves looking for relevant documentation to ensure that their \gls{ai} systems comply with the ethical requirements of existing laws, guidelines, and standards.
Recent studies indicate inadequate model documentation practices, suggesting a gap between \gls{ai} requirements and current practices in model documentation.
To understand this gap and provide actionable guidance to bridge it, we conducted a thematic analysis of 26 guidelines on ethics and \gls{ai}, three \gls{ai} documentation frameworks, three quantitative studies of model cards, and ten actual model cards.
We identified a total of 43 ethical requirements relevant to model documentation and organized them into a taxonomy featuring four themes and twelve sub-themes representing ethical principles.
Our findings indicate that model developers predominantly emphasize model capabilities and reliability in the documentation while overlooking other ethical aspects, such as explainability, user autonomy, and fairness.
This underscores the need for enhanced support in documenting ethical \gls{ai} considerations.
Our taxonomy serves as a foundation for a revised model card framework that holistically addresses ethical \gls{ai} requirements.

\end{abstract}


\section{Introduction}
\label{sec:intro}

The wide adoption of \gls{ai} across professional and private domains has made its responsible development and use a pressing societal topic~\cite{maslej_ai_report_2024}.
\gls{ai} systems mimic human intelligence to autonomously perform complex tasks, offering significant benefits~\cite{sheikh_artificial_2023}.
At the same time, the development of these systems introduces risks, such as consent violations in the collection of large amounts of personal data~\cite{van_der_sype_lawful_2014} and the introduction of biases in \gls{ai}-based decision-making processes~\cite{mehrabi_survey_2022}.
Such risks highlight the need for responsible \gls{ai} engineering~\cite{maalej_tailoring_2023} and regulations to govern \gls{ai} technologies~\cite{european_commission_ai_2024}.
Ethical requirements are, therefore, of crucial importance in complementing technical requirements.

Popular platforms such as Hugging Face and GitHub offer wide access to \gls{ai} models.
On these platforms, \gls{ai} engineers play dual roles: as developers who provide models with certain qualities and as users who integrate the models in their systems and ensure compliance with laws, policies, and standards.
Effective communication between these roles depends on clear documentation of the requirements considered during model development to make informed deployment decisions~\cite{montgomery_empirical_2022}.

The model sharing and deployment platforms promote model cards as central documentation artifacts and boundary objects between model developers and users~\cite{wohlrab_boundary_2019}.
Proposed in 2019, model cards provide structured sections like \textquote{Intended Use} and \textquote{Ethical Considerations} to document model details~\cite{mitchell_model_2019}.
Since then, research in the field of \gls{ai} ethics has evolved significantly~\cite{jobin_global_2019, hagendorff_ethics_2020}, especially with the rise of generative \gls{ai}~\cite{huang_ethical_2025}.
Despite their potential, currently published model cards often lack detailed ethical descriptions~\cite{bhat_aspirations_2023, gao_documenting_2024}.
This shortcoming is possibly due to the initial framework that introduced ethical considerations as a vague concept placed at the end of the document~\cite{mitchell_model_2019}.

Recent work studied the documentation \textit{practices} for \gls{ai} models~\cite{bhat_aspirations_2023, gao_documenting_2024, pepe_how_2024}, but has not used \textit{theoretical}, ethical \gls{ai} frameworks for their analysis.
The objective of this study is to analyze the gap between theory and practice and provide developers and users of \gls{ai} models with holistic guidance on relevant ethical requirements.
We focus on the following research question:

\begin{itemize}[\IEEEsetlabelwidth{iii}]
\item[\textbf{RQ:}] What ethical requirements do \gls{ai} guidelines and documentation frameworks describe in comparison to model cards?
\end{itemize}

To answer this question, we conducted a thematic analysis of \gls{ai} guidelines, \gls{ai} documentation frameworks, and model cards.
The contribution of the study is twofold:

\begin{enumerate}
    \item A taxonomy of 43 ethical requirements relevant for \gls{ai} model documentation, organized into four themes and twelve sub-themes representing key ethical principles.
    \item An analysis revealing the coverage of these requirements within guidelines, documentation frameworks, and model cards.
\end{enumerate}

We discuss how our work helps to bridge the gap between theoretical and practical perspectives by guiding practitioners to identify and refine \gls{ai} requirements and providing researchers with a holistic foundation for future ethical \gls{ai} studies.
We describe our data collection and analysis processes in Section~\ref{sec:method} and the results in Section~\ref{sec:res}.
Then, we discuss the implications of our work in Section~\ref{sec:disc}, study limitations in Section~\ref{sec:threats}, and related work in Section~\ref{sec:rw}. 
Our replication package includes further details on our data collection and analysis~\cite{rep_pack}.

\section{Methodology}
\label{sec:method}

\begin{figure*}
    \centering
    \includegraphics[width=1\linewidth]{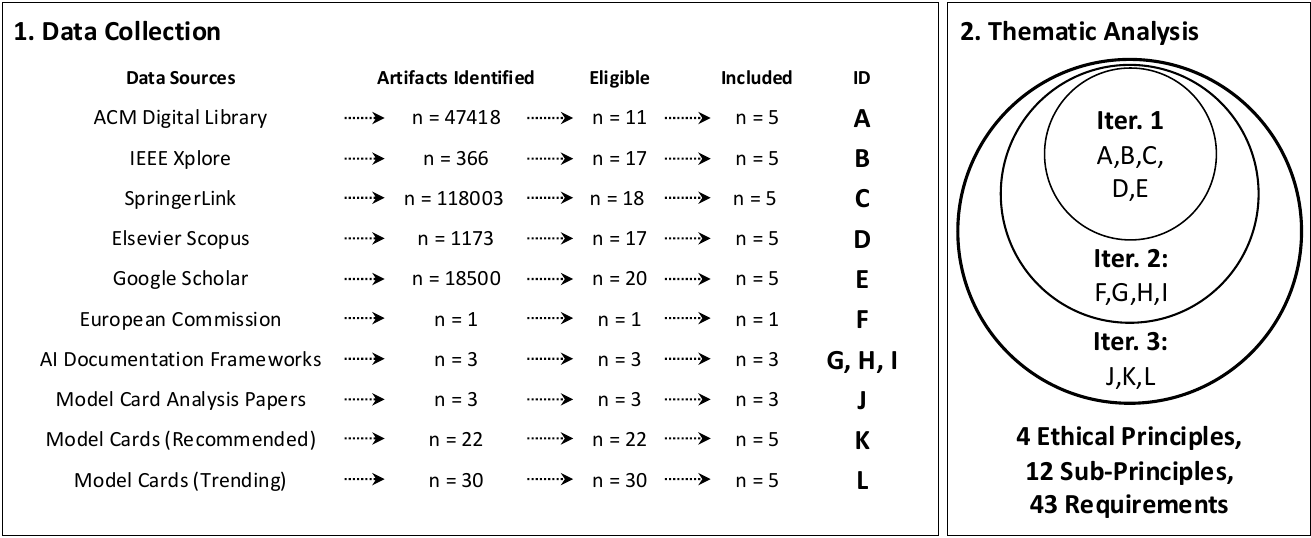}
    \caption{Overview of the study design consisting of two major parts: data collection and thematic analysis.}
    \label{fig:method}
\end{figure*}

To compile ethical requirements for \gls{ai} models and their documentation, the study adopts a two-part methodology encompassing data collection and thematic analysis.
We seek to elucidate the ethical requirements outlined in existing theoretical guidelines and how these are represented within model card implementations.
Figure~\ref{fig:method} provides an overview of the study design, which includes a systematic collection of \gls{ai} guidelines, \gls{ai} documentation frameworks, model card analysis papers, and model cards themselves, as well as the iterative thematic analysis to extract relevant ethical requirements.
All analyzed artifacts are listed in Table~\ref{tab:data}.

\subsection{Collection of AI Guidelines}
\label{sec:method-collection-guide}

To identify relevant \gls{ai} guidelines, we adhered to the comprehensive guidelines proposed by Kitchenham and Charters for conducting \glspl{slr}~\cite{kitchenham_guidelines_2007}.
The methodological rigor of these guidelines is further enhanced by insights from subsequent studies~\cite{brereton_lessons_2007, garousi_guidelines_2019, felderer_guidelines_2020, ralph_paving_2022}.

As data sources, we chose the \textit{ACM Digital Library}, \textit{IEEE Xplore}, and \textit{SpringerLink} due to the indexing of reputable publications in \gls{ai}, software engineering, and requirements engineering. 
Additionally, we added \textit{Elsevier Scopus} and \textit{Google Scholar} to ensure a comprehensive coverage of various databases beyond ACM, IEEE, and Springer.

The search employed the following query that we iteratively refined to maximize the relevance of retrieved artifacts.
The query creation was guided by insights from two significant secondary studies that formed the basis of the data collection process ~\cite{hagendorff_ethics_2020, jobin_global_2019}.
Adjustments included the use of multiple keywords and varied syntactic forms to address terminological diversity: 
\begin{quote}
\centering
(\textit{responsib*} OR \textit{ethic*}) AND\\(\textit{AI} OR \textit{artificial intelligence}) AND \textit{requirement*}
\end{quote}

The search returned 47,418 papers from ACM, 366 from IEEE, 118,003 from Springer, 1,173 from Scopus, and 18,500 from Google Scholar. 
We reviewed the top 100 entries per source with the following inclusion criteria~\cite{felderer_guidelines_2020}:

\begin{enumerate}
\item Published between 2019-01-01 and 2024-09-05\\(to focus on recent studies that also consider new risks introduced by generative \gls{ai})
\item Written in English
\item Peer-reviewed scientific publication
\item Title/abstract refers to \gls{ai}
\item Content covers ethical requirements for \gls{ai}
\end{enumerate}

We conducted a three-stage review process: We screened titles and abstracts in iteration 1 to identify the first 20 potentially relevant papers per database, followed by a full-text analysis in iteration 2 for inclusion decisions.
Iteration 3 involved a \textit{test-retest} verification of the consistency of inclusion decisions~\cite{kitchenham_guidelines_2007}, resulting in the inclusion of two initially excluded papers.
Afterward, we identified and removed duplicates, culminating in a core set of 83 unique papers.
We drew a random sample of five papers per data source. 
This was a trade-off between rigor and coverage based on the assumption that the overlap of ethical requirements in the guidelines would be large.
Our results, summarized in Table~\ref{tab:taxonomy}, reflect this overlap.

Additionally, we included the guidelines \textquote{\gls{altai}} by the European Commission for its holistic perspective and actionable insights derived from over 50 academic and industrial experts.
This guideline contributed ethical requirements that were absent in other guidelines.

\subsection{Collection of AI Documentation Frameworks}
\label{sec:method-collection-frame}

The inclusion of \gls{ai} documentation frameworks in our study was essential as they play a pivotal role in translating theoretical guidelines into practical tools for practitioners, thus bridging the gap between scientific knowledge and practical needs.
Scientists developing such frameworks must cope with trade-offs between simplicity and detail.
The simpler a framework is, the easier it will be accepted by practitioners.
However, important details from theory could then be neglected.
This compromise could lead to widely accepted frameworks that are inadequate or incomplete from a scientific point of view.

Our focus was on following \gls{ai} documentation frameworks:

\begin{itemize}
    \item \textit{Model Cards} by Google~\cite{mitchell_model_2019},
    \item \textit{Fact Sheets} by IBM~\cite{arnold_factsheets_2019},
    \item \textit{Explainability Fact Sheets} by University of Bristol~\cite{sokol_explainability_2020}.
\end{itemize}

To this date, the model card is the only documentation framework focusing on \gls{ai} models.
Due to its adoption by Hugging Face, it became a popular framework among practitioners and was frequently quoted in scientific papers.
Fact sheets consider the whole \gls{ai} system (the model embedded in a system). 
Thus, we expected a strong focus on the perspectives of \gls{ai} users.
Similarly, the explainability fact sheet also considers the \gls{ai} system but with a focus on explainability, which is an essential ethical principle for \gls{ai}.

We excluded other, specialized frameworks such as the \textit{Interactive Model Card} by Crisan et al.~\cite{crisan_interactive_2022} and dataset-focused documentation frameworks~\cite{gebru_datasheets_2021, pushkarna_data_2022, chmielinski_dataset_2022, bender_data_2018} from our data collection, as these frameworks do not focus on ethical requirements of \gls{ai} models.

\subsection{Collection of Model Cards and Related Papers}
\label{sec:method-collection-model}

We performed a triangulated analysis involving (1) recent studies on model card content, (2) model cards highlighted in these studies, and (3) model cards of models trending on Hugging Face.
This approach enabled both an in-depth examination of the primary data and generalizable statements about the content of the model cards.

The considered studies are (1) an analysis of documented ethical considerations by Gao et al.~\cite{gao_documenting_2024}, an analysis of documented datasets, biases, and licenses by Pepe et al.~\cite{pepe_how_2024}, and a general analysis of model card content by Bhat et al.~\cite{bhat_aspirations_2023}.
None of them referred to \gls{ai} guidelines (i.e.,~theory) as a comparison to the model card content (i.e.,~practice).
Only reassessing the results of those studies will likely lead to an incomplete observation of model cards compared to what they are supposed to document.
Therefore, we randomly sampled five out of 22 model cards positively highlighted in the three studies and supplemented them with five random samples from the 30 most trending models on Hugging Face in January/February 2025. 
This selection represents the range of models that either have comprehensive ethical documentation or whose documentation is sufficient to appeal to a broad spectrum of users.

\subsection{Thematic Analysis}
\label{sec:method-analysis}

The thematic analysis sought to structure and interpret the ethical \gls{ai} requirements identified within the collected artifacts.
Adhering to the guidelines by Braun and Clarke~\cite{braun_using_2006}, our analysis spanned three comprehensive iterations, each involving five key decisions and six steps for analysis.
We slightly adapted the analysis decisions per iteration and report the ones made for the final iteration in the following.

(1) Definition of themes:
We defined themes and sub-themes as ethical principles encompassing codes that represent ethical requirements of model users.
Themes, sub-themes, and codes should form a hierarchical taxonomy.
We added a user-centered question to each sub-theme to make them more tangible and avoid ambiguity, for example: \textit{\textquote{Accuracy: How do developers ensure continuously correct model output?}}.
The codes are formulated as requirements for the \gls{ai} model documentation.
Their implementation answers the respective sub-theme question, for example, \textit{\textquote{(R02) Explain mechanisms in place to continuously ensure accuracy}}.

(2) Analytical scope:
We limited the analysis to content applicable to \gls{ai} models.
Aspects only relevant to the broader scope of \gls{ai} systems, such as accessibility and universal design, were excluded to maintain focus on model requirements.

(3) Analytical approach:
We adopted a deductive analytical approach.
Our analysis was driven by a specific research question with the aim of fitting relevant data into a requirements-focused coding frame.

(4) Emergence of themes:
We mostly applied a semantic level of coding by structuring and formulating our codes and sub-themes based on the content of the analyzed guidelines.
However, their integration into main themes required the interpretation of latent concepts, as the guidelines sometimes used contradictory or overlapping structures.

(5) Epistemological stance:
We chose the constructionist paradigm and recognized ethical requirements as social constructs for the interaction between model developers and users.

All three time-intensive iterations followed the six phases of thematic analysis:
(1) familiarize yourself with the data,
(2) generate initial codes,
(3) search for themes,
(4) review themes,
(5) define and name themes, and
(6) produce the report.
Per iteration, we increased the number of analyzed artifacts (see Figure~\ref{fig:method}).
Each iteration was mainly performed by one author to analyze the large amount of content in-depth.
Phase~5 was accompanied by a second author to review the clarity and relevance of all codes and themes.
The open-source tool \textit{QualCoder} facilitated the analysis process~\cite{ccbogel_qualcoder_nodate}.
We detail the three iterations in the following.

(Iteration 1)
The first iteration was led by the second author and supported by the first author.
We analyzed the 25 scientific guidelines and created a preliminary structure with ten themes, each encompassing several codes. 

(Iteration 2)
The first author led the second iteration, assisted by the third author.
We re-evaluated all data from the first iteration and deconstructed the codes into smaller components, which helped to clarify and reduce the number of themes.
We extended the analysis to the \gls{altai} guidelines and the three documentation frameworks.
The result was five main themes.

(Iteration 3)
The final iteration was again conducted by the first author, assisted by the third author.
It included all artifacts, including the model card analysis papers and model cards.
We decided that each previous code should become a sub-theme corresponding to an ethical principle.
The new codes should each focus on a unique ethical consideration that is part of the model documentation.
We formulated these codes as functional requirements.
This last iteration further clarified the taxonomy and helped to eliminate overlaps between the themes and strengthen their cohesion.

\section{Results}
\label{sec:res}

\input{tables/data}

\input{tables/taxonomy}

The thematic analysis resulted in a taxonomy featuring four central themes representing the ethical principles \textit{Reliability}, \textit{Transparency}, \textit{Empowerment}, and \textit{Beneficence}. 
Each principle includes three sub-principles, further broken down into various codes, representing requirements. 
Table~\ref{tab:taxonomy} provides an overview of the entire taxonomy in relation to the data sources.
The values in the table cells indicate the presence (\textit{1}) or absence (\textit{0} or \textit{empty cell}) of a requirement in the analyzed artifact.
For clarity, we consolidated the \gls{ai} guidelines into six columns relative to their data source, such as ACM or IEEE.
Similarly, we grouped model card analysis papers, recommended model cards, and trending model cards into one column each.
Consequently, for columns A-E and J-L, the matrix values represent the presence totals.
This approach was particularly helpful in providing an overview of the requirements included in the model card analysis papers, as each of the three studies had a different substantive focus, for instance, Pepe et al.~on datasets, biases, and licenses~\cite{pepe_how_2024}.
We describe each taxonomy component in the following.

\subsection{Reliability}
\label{sec:res-reliability}

Reliable \gls{ai} models should function as expected even when presented with novel input~\cite{eu_2020}.
Such models act consistently, are dependable, and align with intended purposes~\cite{sanderson_towards_2022}.
The sub-principles of Reliability include \textit{Accuracy}, \textit{Safety and Security}, and \textit{Auditability}.

\vspace{6pt}
\noindent
\textit{\textbf{Accuracy: How do the developers ensure continuously correct model output?}}
Accuracy involves comparing predicted model output with observed evidence~\cite{giudici_artificial_2024}.
An accurate model can correctly identify patterns and classify information~\cite{lopez_requirements_2024}.
This sub-principle comprises the following three requirements.

\textbf{(R01) Explain quality of training/evaluation data, process, and results} is a dominant topic in the guidelines, documentation frameworks, and model cards.
Guidelines mandate that model developers ensure data is current, of high quality, complete, representative of its environment~\cite{eu_2020}, and tested~\cite{lu_towards_2022, giudici_artificial_2024}.
Any testing processes should be documented transparently~\cite{eu_2020}.
All frameworks demand explanations of data~\cite{sokol_explainability_2020}, model training configurations such as hyperparameters~\cite{arnold_factsheets_2019}, and performance measures~\cite{mitchell_model_2019}.
Such information is consistently available across the analyzed model cards.

\textbf{(R02) Explain mechanisms in place to continuously ensure accuracy} is less visible in frameworks and model cards.
While guidelines require ongoing accuracy monitoring~\cite{jobin_global_2019}, including human oversight~\cite{diaz-rodriguez_connecting_2023} and retraining criteria to counter data drifts~\cite{lu_towards_2022, andersen_design_2024}, only the fact sheets framework mentions such mechanisms~\cite{arnold_factsheets_2019}.

\textbf{(R03) Explain factors that influence model accuracy} is mentioned by two guidelines that demand insights into model input features affecting accuracy~\cite{sanderson_towards_2022, lopez_requirements_2024}.
Although this is an explicit section in the model card framework~\cite{mitchell_model_2019}, practical descriptions of factors seem scarce.
Only Bhat et al.~touch upon target distributions, which represent input that a model can classify correctly~\cite{bhat_aspirations_2023}.
Such descriptions are, however, generally brief.

\vspace{6pt}
\noindent
\textit{\textbf{Safety and Security: How do the developers ensure protection of individuals and groups from harm?}}
A safe \gls{ai} model should act robust and error-free, while a secure model is protected from external threats~\cite{lopez_requirements_2024, eu_2020}.
Models and development processes with privacy protection \cite{van_der_sype_lawful_2014} are safeguarded against attacks leading to sensitive data leaks~\cite{khalid_privacy-preserving_2023} and respect the right of an individual to exist and act without undesired public scrutiny~\cite{slavkovik_mythical_2023}.
This sub-principle encompasses the following six requirements.

\textbf{(R04) Explain security measures for mitigating external threats}, \textbf{(R05) Explain safety measures for increasing robustness and mitigating harm}, and \textbf{(R06) Explain measures for protecting the privacy of data subjects} are frequently highlighted in guidelines, partially in frameworks, but rarely in model cards.
Notable security measures include data minimization~\cite{keskinbora_medical_2019} and protection against adversarial attacks~\cite{akula_ethical_2021}, along with human-focused measures like data access control~\cite{khalid_privacy-preserving_2023} and risk training~\cite{eu_2020}.
Frameworks mention threats such as information leakage and request data security information~\cite{sokol_explainability_2020, arnold_factsheets_2019}.
Guidelines state safety measures like implementing fallback plans for unexpected inputs~\cite{eu_2020} and testing within an \textit{ethical sandbox}~\cite{lu_responsible-ai-by-design_2023}, while frameworks request risk mitigation strategies~\cite{mitchell_model_2019} and mechanisms alerting users about incompatible inputs~\cite{arnold_factsheets_2019}.
Notable privacy protection measures include federated learning~\cite{diaz-rodriguez_connecting_2023}, data encryption~\cite{ryan_artificial_2021}, and differential privacy~\cite{lu_towards_2022}.
On the practical side, only safety measures are discussed by a few model cards, for example, how the model was tested for harmful output~\cite{salesforce_salesforcectrl_nodate}.

\textbf{(R07) Explain duration of safety and security coverage} is a nuanced requirement found in a single guideline~\cite{eu_2020}.
It relates to model users who need assurance regarding long-term model maintenance, particularly in the face of new threats.

\textbf{(R08) Provide risk assessment for model usage} is more prevalent than the previous requirement.
Guidelines refer to regulations concerning high-risk \gls{ai} applications~\cite{iliadis_ai-gfa_2024} and urge assessments of safety and security risks that might harm individuals~\cite{eu_2020}.
Frameworks also require risk assessments.
Model cards, however, largely omit detailed discussions of risks.
For example, almost all recommended model cards discuss biases and performance limitations but lack comprehensive risk assessments.

\textbf{(R09) Provide recommendations for users to mitigate risks} appear only in the model card framework and recommended model cards.
The framework requests additional recommendations for model use and testing~\cite{mitchell_model_2019}.
The model cards suggest that users carry out further risk assessments~\cite{salesforce_salesforcectrl_nodate} or fine-tune the model for specific cases~\cite{technology_innovation_institute_tiiuaefalcon-180b_nodate}.

\vspace{6pt}
\noindent
\textit{\textbf{Auditability: How do the developers ensure that the model qualities are reviewed and verified?}}
An \gls{ai} model should have mechanisms in place for regular assessments to verify compliance with laws, policies, or specifications~\cite{diaz-rodriguez_connecting_2023, eu_2020}.
Auditability consists of three requirements.

\textbf{(R10) Explain measures in place to ensure auditability} especially concerns the reproducibility of model behavior~\cite{diaz-rodriguez_connecting_2023, lier_what_2024, kuwajima_adapting_2019, eu_2020} and involves integrating evaluation and monitoring stages in the model life cycle~\cite{lopez_requirements_2024}. Neither the frameworks nor the models provide related information.

\textbf{(R11) Provide details on audit procedures and results} focuses on periodic external assessments measuring criteria like accuracy and fairness~\cite{zhang_trusted_2021, akhtar_transparency_2024}.
Frameworks demand listings of third-party audits and results~\cite{arnold_factsheets_2019}.
Among the models, only one reports on audits~\cite{salesforce_salesforcectrl_nodate}.

\textbf{(R12) Provide certification to demonstrate compliance with laws, policies, and standards} is mentioned in multiple guidelines~\cite{leopold_mastering_2023, jobin_global_2019, eu_2020}.
For example, one guideline introduced the concept of \textit{ethical credentials} to enable \gls{ai} in riskier use cases~\cite{lu_responsible-ai-by-design_2023}.
No framework or model features such certification information.

\subsection{Transparency}
\label{sec:res-trans}

\textit{Transparency} means that the \gls{ai} capabilities and decision-making processes are presented clearly and accessibly to all stakeholders~\cite{lopez_requirements_2024}.
Its sub-principles are \textit{Communication of Capabilities}, \textit{Explainability}, and \textit{Traceability}.

\vspace{6pt}
\noindent
\textit{\textbf{Communication of Capabilities: How do the developers ensure to communicate what the model can and cannot do?}}
This sub-principle involves open communication of basic \gls{ai}-related information like the purpose of the model and the availability of training data.
This was the most common sub-principle in the model cards.

\textbf{(T01) Explain intended uses and applicable domains} and \textbf{(T02) Explain out-of-scope uses and model limitations} are about essential information that potential model users need to make informed usage decisions~\cite{ryan_artificial_2021}.
The model card framework provides sections dedicated to these requirements~\cite{mitchell_model_2019}.
Nevertheless, while intended uses are documented in almost all model cards, fewer discuss limitations, such as supported languages~\cite{technology_innovation_institute_tiiuaefalcon-180b_nodate}, or inappropriate uses.
Inappropriate uses could be, for example, fact-checking~\cite{black_forest_labs_black-forest-labsflux1-dev_nodate} or autonomous decision-making~\cite{openai_openaiwhisper-medium_nodate}.

\textbf{(T03) Provide details on model architecture and algorithms}, \textbf{(T04) Provide training and evaluation data}, and \textbf{(T05) Explain trade-offs made for competing design goals} demand greater detail on model capabilities and limitations.
Model cards report applied hyperparameters~\cite{nomic_ai_nomic-ainomic-embed-text-v2-moe_nodate} and fine-tuning processes~\cite{technology_innovation_institute_tiiuaefalcon-180b_nodate}.
Guidelines and frameworks request evaluation data to enable replication of claimed performance~\cite{mitchell_model_2019}.
Guidelines often discuss trade-offs, like utility versus privacy or accuracy versus fairness~\cite{iliadis_ai-gfa_2024}, but model cards never do.

\textbf{(T06) Explain how users should correctly interact with the model} is rarely mentioned in guidelines but is frequently available in model cards, often through code examples.

\textbf{(T07) Refer to additional documentation} is not an ethical requirement, but is relevant to describe the current practice as large organizations provide additional information through various platforms, including their websites. 
All model cards refer to such additional documentation.

\vspace{6pt}
\noindent
\textit{\textbf{Explainability: How do the developers ensure interpretable model predictions?}}
Explainability includes two requirements and demands the clarification of algorithmically generated decisions with justifications that are also comprehensible to non-\gls{ai} experts~\cite{slavkovik_mythical_2023, eu_2020}.
However, none of the model cards include any information on explainability.

\textbf{(T08) Explain how and for whom the model generates explanations} is noted as a requirement in nearly all guidelines and frameworks, emphasizing users' right to information about reasoning processes~\cite{lopez_requirements_2024}.

\textbf{(T09) Explain how explanation quality was tested} seeks quality criteria for explanations.
For instance, Sokol and Flach specify criteria such as soundness and contextfulness~\cite{sokol_explainability_2020}.

\vspace{6pt}
\noindent
\textit{\textbf{Traceability: How do developers ensure documented and reproducible model development and predictions?}}
The three traceability requirements demand a complete account of the data provenance, development processes, and artifacts related to \gls{ai} decisions~\cite{slavkovik_mythical_2023}.

\textbf{(T10) Explain where, how, and when training data was obtained and processed} is about tracing data sources and modifications.
Related information is absent from frameworks and model cards.

\textbf{(T11) Explain how model and data modifications are logged} refers to logging mechanisms, like a version-based repository~\cite{lu_towards_2022, lu_responsible-ai-by-design_2023}. 
While advanced concepts like a \textit{\gls{sdoc}}~\cite{zhang_trusted_2021} and \textit{\gls{bom}}~\cite{lu_towards_2022} have been adapted in some software engineering domains, analyzed model cards only provide basic version numbers and modification dates~\cite{bhat_aspirations_2023}.

\textbf{(T12) Explain how user interaction data is logged} holds relevance when models send usage data to developers, for example, to monitor data drifts or errors~\cite{ryan_artificial_2021, eu_2020}.
This applies not only to hosted models but also to self-deployed ones that contain any form of backdoor.
However, related information is absent in frameworks and model cards. 

\subsection{Empowerment}
\label{sec:res-empow}

\textit{Empowerment} focuses on strengthening human rights, especially the rights of model users and individuals affected by \gls{ai} predictions~\cite{jobin_global_2019}.
Developers must be accountable for development processes and released models~\cite{diaz-rodriguez_connecting_2023}.
Sub-principles are \textit{User Autonomy}, \textit{Consent and Control}, and \textit{Liability}.

\vspace{6pt}
\noindent
\textit{\textbf{User Autonomy: How do the developers ensure empowering rather than restricting or manipulating users?}}
User autonomy addresses the human decision-making capacity~\cite{lopez_requirements_2024}.
\gls{ai} models should act as enablers~\cite{eu_2020}, and users should have control over decision-making~\cite{ryan_artificial_2021}.
Although user autonomy is a central ethical principle, information on its three requirements can only be found in the guidelines.

\textbf{(E01) Explain how the model informs users about interacting with an AI} is crucial when models simulate human conversations~\cite{ryan_artificial_2021, eu_2020}, especially when the decisions impact individuals~\cite{lopez_requirements_2024}.

\textbf{(E02) Explain how the model assists users in decision-making without manipulation} requests enhanced user control by requiring the \gls{ai} to offer meaningful choices and not limit a decision to a single option~\cite{jedlickova_ethical_2024}. 

\textbf{(E03) Explain how human oversight can influence or override model predictions} mandates human intervention possibilities~\cite{lopez_requirements_2024, lu_responsible-ai-by-design_2023}, for example, with models that react and adjust their responses to external interruptions~\cite{liu_enhancing_2024}.

\vspace{6pt}
\noindent
\textit{\textbf{Consent and Control: How do the developers ensure the authority of data subjects?}}
Data subjects are all individuals and organizations whose data is used to develop the model.
They have a right to privacy and control over their data~\cite{akhtar_transparency_2024}.

\textbf{(E04) Explain how data subjects are informed of data processing and can manage consent} is crucial, especially when the amount of data used for model training increases.
Guidelines emphasize that data subjects must be able to control their consent to data use~\cite{kuwajima_adapting_2019}, including the possibility of withdrawal~\cite{eu_2020}.

\textbf{(E05) Explain how data subjects can access and request the change or deletion of their data} urges developers to provide insights and rectification opportunities to data subjects ~\cite{lopez_requirements_2024}.
Among the frameworks, only fact sheets mention related information, and one model card highlights compliance with applicable laws and regulations in data processing~\cite{gao_documenting_2024}.

\vspace{6pt}
\noindent
\textit{\textbf{Liability: How do the developers ensure their accountability for the model?}}
If an \gls{ai} harms individuals, for example, due to unfair model predictions, responsible actors should be held legally accountable~\cite{golbin_responsible_2020}.

\textbf{(E06) Name responsible and liable entities throughout the model lifecycle} is an essential measure for accountability indication.
Despite Hugging Face showing a connection between a model and a developer account, it does not guarantee the disclosure of the responsible legal entity.
Bhat et al. identify contact information in model cards, while we only found one model card that directly provides contact information of a developer group~\cite{agentica_agentica-orgdeepscaler-15b-preview_nodate}.

\textbf{(E07) Explain procedures for contesting and redressing harmful model behavior} and \textbf{(E08) Explain procedures for handling liability claims} require redressing mechanisms and procedures for unjust behavior, but related information is not available in the analyzed model cards.
Only Gao et al. point out some model cards where developers encourage users to report ethical concerns~\cite{gao_documenting_2024}.

\textbf{(E09) Explain how model developers report harm to affected parties and the public} addresses the need for transparency and accountability after incidents.
Guidelines refer to this as \textquote{acting with integrity}~\cite{ryan_artificial_2021}.
Relevant information is absent from the model cards.

\textbf{(E10) Explain which responsibilities the model developers disclaim}, has higher attention in practice.
The model card framework and analysis papers describe license declarations~\cite{mitchell_model_2019, bhat_aspirations_2023} or disclaimers regarding the responsibility for model predictions~\cite{gao_documenting_2024} or unintended use cases~\cite{pepe_how_2024}.

\subsection{Beneficence}
\label{sec:res-bene}

Beneficence embodies that \gls{ai} technology should benefit all humanity~\cite{keskinbora_medical_2019}, aligning, for example, with the \textit{Sustainable Development Goals} established by the United Nations~\cite{diaz-rodriguez_connecting_2023}. 
Related terms are social and common good~\cite{ryan_artificial_2021}.
Included sub-principles are \textit{Fairness}, \textit{Social Beneficence}, and \textit{Environmental Beneficence}.

\vspace{6pt}
\noindent
\textit{\textbf{Fairness: How do the developers ensure non-discriminatory, equitable model output?}}
Fairness mandates equality in the treatment of individuals and groups~\cite{eu_2020}, avoiding biases and discriminatory model outputs~\cite{sanderson_towards_2022, giudici_artificial_2024}.
Despite extensive fairness discussions in guidelines and explicit questions in the frameworks, related information is nearly non-existent in model cards.

\textbf{(B01) Explain how diverse stakeholders were involved in model development} emphasizes the involvement of stakeholders who are affected by the model output; either directly in the development process or through feedback~\cite{iliadis_ai-gfa_2024}, which is termed \textit{society-in-the-loop}~\cite{liu_enhancing_2024} or \textit{participatory design}~\cite{diaz-rodriguez_connecting_2023}.
Implementing this requirement reduces power imbalances between developers and other stakeholders~\cite{jedlickova_ethical_2024}.

\textbf{(B02/03/04) Explain bias detection techniques applied to datasets (pre-processing), during model training (in-processing), and to model output (post-processing).} 
These requirements refer to three areas in which biases must be analyzed and removed; specifically, through pre-processing steps like demographic balancing~\cite{diaz-rodriguez_connecting_2023} and augmenting missing data~\cite{lu_towards_2022}, in-processing procedures like adversarial learning~\cite{zhang_trusted_2021}, and post-processing measures such as fairness-focused testing~\cite{arnold_factsheets_2019} and sanity checks~\cite{diaz-rodriguez_connecting_2023}.
However, model cards provide only minimal information on pre-processing and note potential model biases without details on bias detection or mitigation.

\vspace{6pt}
\noindent
\textit{\textbf{Social Beneficence: How do the developers ensure positive impact of the model on the well-being of society?}}
This sub-principle focuses on promoting societal welfare through the use of \gls{ai}~\cite{iliadis_ai-gfa_2024}.

\textbf{(B05) Explain positive impact the model can have on individuals and society} is a requirement that is not featured by any framework.
Still, a few model cards emphasize potential use cases such as misinformation detection~\cite{salesforce_salesforcectrl_nodate}, accessibility tool development~\cite{openai_openaiwhisper-medium_nodate}, and enabling research projects~\cite{gao_documenting_2024}.

\textbf{(B06) Explain usage permissions and restrictions} focuses on strengthening intended uses and restricting malicious ones.
Interestingly, guidelines rarely discuss this topic.
A single guideline mentions unintended dual uses, such as for military applications~\cite{lier_what_2024}.
Frameworks and model cards mention this requirement mostly in connection with license declarations.
One model card highlights the role of permissive licenses in the democratization \gls{ai} technology~\cite{agentica_agentica-orgdeepscaler-15b-preview_nodate}.
However, such licenses do not limit the unintended use cases.

\vspace{6pt}
\noindent
\textit{\textbf{Environmental Beneficence: How do the developers ensure environmental sustainability?}}
Environmentally sustainable models and development processes are linked to advancing societal good~\cite{diaz-rodriguez_connecting_2023, pham_shapere_2020, pham_role_2021}.
However, this information is often lacking, even in frameworks, and is only sporadically included in model cards.

\textbf{(B07) Explain positive impact the model can have on the environment} underscores sustainable model use cases that counteract ecological challenges, in particular climate change~\cite{diaz-rodriguez_connecting_2023} and shrinking biodiversity~\cite{jobin_global_2019}.

\textbf{(B08) Provide measured values for environmental impact metrics} calls for reporting metrics like carbon footprint, energy consumption, and training efficiency~\cite{diaz-rodriguez_connecting_2023, eu_2020}.

\textbf{(B09) Explain efforts made to minimize the environmental footprint of the model} requests explicit steps toward sustainable development, such as data reduction, model compression, multi-use case adaptation, and using renewable energies~\cite{diaz-rodriguez_connecting_2023}.
Only one model card details cost-reduction measures~\cite{agentica_agentica-orgdeepscaler-15b-preview_nodate}, while Gao et al. found carbon footprint reports in 15 model cards~\cite{gao_documenting_2024}.

\section{Discussion}
\label{sec:disc}

\subsection{The Gap Between Theory and Practice of Ethical AI}
\label{sec:res-gap}

Our analysis distilled a wide range of ethical \gls{ai} requirements: from theory (ethical \gls{ai} guidelines and \gls{ai} documentation frameworks) and practice (model cards).
We observed a divergence between theory and practice, characterized by a decreasing number of requirements from theory to practice: The guidelines discuss almost all requirements, the frameworks focus on a subset, and the model cards only provide information for a few of those requirements.

Our results signal that the three \gls{ai} documentation frameworks studied do not capture the latest advances in research on responsible \gls{ai} requirements and that model developers either lack guidance or motivation to fully implement these frameworks.
The analyzed model cards seem to particularly focus on communicating model capabilities and reporting certain reliability measures to highlight advantages over competitors.
Limitations and risks receive rather a superficial acknowledgment, for example, by including notes and disclaimers regarding potential model biases.
Profound ethical reflections and how to operationalize and address them (e.g., by model users) are rare and incomplete.

Overall, this study points to a gap between rigorous research in ethics and \gls{ai} on the one hand and documentation practices that focus on groundbreaking technical \gls{ai} capabilities but rather neglect reflections on responsible engineering processes on the other hand.

\subsection{Closing the Gap}
\label{sec:disc-close}

We think that it is crucial to further understand and address the gap between theory and practice in considering ethical \gls{ai} requirements. 
Multiple legislators like the European Union, the Chinese Government, or the Australian Government have already established regulations that mandate ethical assessments of \gls{ai} systems~\cite{european_commission_ai_2024, austrailia-ai-2024, china-ai-2024}.
Therefore, potential users of \gls{ai} models need specific information about those models, their development, and maintenance, as well as guidance on how to address potential risks and limitations when using a model.

Model cards have emerged as central boundary objects between model developers and users.
In current practice, model developers may perceive model cards merely as marketing tools to showcase model capabilities.
We suggest that model cards should evolve into standardized requirements documents enriched with a fine-grained structure.
This structure can guide model developers in considering and documenting ethical requirements and help model users not only recognize relevant requirements and risks but also make informed decisions about how to use the models and how to address those requirements and risks.
Such structured information could also be machine-readable, making model documentation similar to frameworks such as \gls{sdoc}~\cite{zhang_trusted_2021} and \gls{bom}~\cite{lu_responsible-ai-by-design_2023}, which transparently depict the supply chain behind a product.

Our taxonomy of ethical principles and requirements can inform the advancement of the model card framework, serving as a guide for various stakeholders, including model developers, users, auditors, hosting platforms, and researchers: 
\begin{itemize}
    \item Model developers can utilize the taxonomy to understand, manage, and evaluate relevant ethical requirements during model development and maintenance.
    \item Model auditors may use the taxonomy as a \textquote{checklist} to ensure that model cards are adequately informative and give precise recommendations on what to improve and how.
    \item Model users can use the taxonomy to guide their elicitation and implementation of ethical \gls{ai} requirements for their own \gls{ai} systems. 
    \item Platforms such as Hugging Face may adopt our taxonomy to enhance their original model card strategy and promote responsible \gls{ai} engineering practices.
    \item Researchers may immerse themselves in the single sub-principles and requirements to better understand the underlying challenges and how to address them.
\end{itemize}

These measures can shift model card documentation from merely presenting model capabilities and superficially naming risks toward delivering a holistic requirements elicitation and validation framework for responsible  \gls{ai} engineering.

\subsection{Further Analyzing the Gap}
\label{sec:disc-analyze}

While previous studies examined the \textquote{as-is} state of model card documentation~\cite{gao_documenting_2024, pepe_how_2024, bhat_aspirations_2023}, our work complements these studies with a \textquote{should-be} perspective rooted in the guidelines and documentation frameworks.
We have developed a taxonomy that enables a foundational comparison between theoretical and practical perspectives on ethical requirements for \gls{ai} models.
Future research endeavors should build on this taxonomy by conducting quantitative analyses of model documentation to provide additional insights into the landscape of ethical documentation in practice.
However, previous studies and ours have already indicated a persistent trend that model cards inadequately document ethical considerations.
Ultimately, this means that the majority of the analyzed artifacts will very likely provide no or only superficially relevant information.
Nevertheless, such studies can provide researchers and tool developers with further data to detect low-quality or missing documentation sections.
This could promote documentation tools such as \textit{DocML} by Bhat et al.~\cite{bhat_aspirations_2023}.

Our findings also suggest that model developers may experience various barriers regarding the use of the original model card framework or the consideration of ethical requirements, including a lack of awareness, comprehension challenges, disinterest, or perceived irrelevance.
Understanding how practitioners perceive and utilize ethics guidelines is fundamental to explaining the gap between theory and practice. 
Exploring their perspectives fosters better alignment between theoretical frameworks and practical acceptance and paves the way for meaningful advancements in responsible \gls{ai} engineering.

\section{Threats to Validity}
\label{sec:threats}

\subsection{Validity of the Data Collection}
\label{sec:threats-collection}

Our findings are impacted by specific design decisions made during the data collection process, particularly concerning the \gls{ai} guidelines and model cards. 
We identified 83 relevant guidelines that were eligible for this study, but acknowledge that the true number of such guidelines is likely higher.
Furthermore, we conducted an in-depth analysis only on a random sample of 25 of these guidelines. 
The rationale for this sampling decision and the quality measures taken are detailed in Section~\ref{sec:method-collection-guide}.
We estimate that significant new insights from additional studies are unlikely, given the comprehensiveness of our sample.

The analyzed guidelines refer to \gls{ai} systems in general and not only to \gls{ai} models.
However, as elaborated in Section~\ref{sec:method-analysis}, we assessed models as integral components of \gls{ai} systems and assumed that model-related requirements are a subset of system requirements.
Therefore, we do not consider the choice of system-focused guidelines as a threat to validity.

Regarding model card data, we balanced between direct primary data analysis and leveraging secondary data from three related studies~\cite{gao_documenting_2024, bhat_aspirations_2023, pepe_how_2024}, as we describe in Section~\ref{sec:method-analysis}.
There is a risk that the secondary data may not have the specificity required for our analysis.
However, the three studies provide comprehensive insights, particularly concerning ethical considerations.
Our taxonomy, detailed in Table~\ref{tab:taxonomy}, enables a comparative analysis of secondary data (column J) against primary data (column K).
We found no additional insights from the primary data beyond those already extracted from secondary data.
Consequently, we consider our conclusions drawn from the primary and secondary data to be valid and generalizable.

\subsection{Validity of the Thematic Analysis}
\label{sec:threats-analysis}

The development of our themes, sub-themes, and codes was an iterative and reflective process. 
However, the resulting codes, representing model card requirements, are not uniform in granularity. 
For instance, a specific requirement like \textquote{Explains duration of safety and security coverage} contrasts with the broader \textquote{Explains security measures for mitigating external threats.}
While the former demands precise temporal information, the latter encompasses a set of potential security measures.
Nevertheless, uniform granularity was not the objective for the taxonomy.
Our primary goal was to develop a taxonomy in which each code possesses a single, unique, and relevant purpose.

The creation of the taxonomy was influenced by our subjective perspectives.
However, we adhered to a systematic thematic analysis process~\cite{braun_using_2006}, with analysis decisions and quality measures outlined in Section~\ref{sec:method-analysis}.
Our taxonomy is a product of current, relevant information from theory and practice.
We are aware that this information is subject to change.
Consequently, regular taxonomy reviews and updates are necessary to maintain its relevance.
We also acknowledge that other researchers might structure ethical \gls{ai} requirements differently.
For example, we considered aligning our taxonomy with the \gls{altai} guidelines by the European Commission~\cite{eu_2020}, which represents the most holistic and actionable artifact we analyzed.
However, we assessed \gls{altai} to be incomplete (as indicated in column F of Table~\ref{tab:taxonomy}) and consider its first theme, \textit{Human Agency and Oversight}, to be an artificial construct with overlaps to other themes.
Additionally, while the \gls{altai} guidelines reflect a European viewpoint, our taxonomy captures a more global perspective with various data sources from theory and practice.

\section{Related Work}
\label{sec:rw}

\subsection{Documentation Frameworks}
\label{sec:rw-doc}

The landscape of documentation frameworks in technical domains is diverse, serving crucial roles in conveying essential information between developers.
Developers have leveraged structured documentation artifacts such as reference documentation of \glspl{api}, \textit{README} files, and computational notebooks with code and documentation blocks to succinctly document software.
Several studies have critiqued the insufficient quality of such documentation, highlighting the necessity for improvements~\cite{maalej_patterns_2013, uddin_how_2015, puhlfuerss_exploratory_2022, pimentel_large-scale_2019, aghajani_software_2019}. 
Also, efforts have been made to address these limitations and provide guidance to improve documentation practices~\cite{kruse_can_2024, nassif_evaluating_2025, wang_documentation_2022}.

Various documentation frameworks have also emerged for \gls{ai} systems, models, and datasets.
Frameworks such as fact sheets~\cite{arnold_factsheets_2019} and explainability fact sheets~\cite{sokol_explainability_2020} concentrate on system-level documentation, while model cards~\cite{mitchell_model_2019} and their interactive model card variant~\cite{crisan_interactive_2022} focus on the models themselves.
Data-centric frameworks include data statements~\cite{bender_data_2018}, dataset nutrition labels~\cite{chmielinski_dataset_2022}, data cards~\cite{pushkarna_data_2022}, and datasheets~\cite{gebru_datasheets_2021}, some of which, unlike the model card framework, pose more explicit ethical questions to the documentation authors.

Model cards, originally introduced by Mitchell et al.~\cite{mitchell_model_2019}, aim to encapsulate key information about \gls{ai} models.
Despite their potential in promoting responsible \gls{ai} engineering, initial empirical studies, such as those by Bhat et al.~\cite{bhat_aspirations_2023}, reveal a general neglect of ethical aspects in practice.
Subsequent studies by Pepe et al.~\cite{pepe_how_2024} and Gao et al.~\cite{gao_documenting_2024} further highlight deficiencies in documenting ethical considerations.
These studies primarily present an \textquote{as-is} analysis of current documentation practices.
We complement this research by analyzing the ethical principles and requirements discussed in more theoretical guidelines and documentation frameworks.
We then examined the extent to which these principles and requirements are mentioned in practice.

\subsection{Ethical Requirements for AI}
\label{sec:rw-req}

The articulation of ethical considerations for \gls{ai} is well-established, with comprehensive guidelines published by academic, industrial, and public sectors.
Already before the emergence of large language models, Jobin et al.~\cite{jobin_global_2019} and Hagendorff~\cite{hagendorff_ethics_2020} cataloged the breadth of ethical principles articulated across the field.
The discourse incorporates themes such as responsibility, trustworthiness, human-centeredness, and social good, which we unify under the umbrella of \textit{Ethics} for clarity in our study~\cite{vassilakopoulou_responsible_2022}.

The practical implementation of these ethical principles, however, poses significant challenges.
Theodorou and Dignum~\cite{theodorou_towards_2020} argued that transitioning from theory to actionable requirements in \gls{ai} systems is complex but urgently needed.
Similarly, Maalej et al.~discussed how to operationalize responsible \gls{ai} in practice by tailoring established requirements engineering approaches for \gls{ai} systems~\cite{maalej_tailoring_2023}.
Our study synthesizes existing ethical guidelines into a cohesive taxonomy of requirements for \gls{ai} model documentation, facilitating not only the integration of ethical considerations into such documentation but also the identification and implementation of those requirements in the final \gls{ai} systems.

\section{Conclusion}
\label{sec:conc}

Platforms like Hugging Face have simplified the use of \gls{ai} models, leading to increased adoption across various domains.
This leap in popularity necessitates an emphasis on ethical requirements due to emerging risks such as data leakage, safety issues, and unfair or discriminatory \gls{ai} behavior.
However, comprehensive documentation of ethical requirements remains rare in practice.
Recent studies indicate that model developers often do not adequately discuss ethical considerations in their model cards~\cite{bhat_aspirations_2023}.

For a full picture, we extend current research by analyzing and comparing ethical requirements described in \gls{ai} ethics guidelines and documentation frameworks with those shared in model cards.
The analysis resulted in a taxonomy of 43 requirements organized into four ethical principles and twelve sub-principles.
Furthermore, we identified a critical gap between the ethical requirements demanded by the guidelines and the information provided in model cards.

The findings underscore the need for clearer guidance from researchers and \gls{ai} platforms to support practitioners in recognizing and addressing the full scope of ethical requirements discussed in the literature.
Our study marks a step towards bridging this gap between theory and practice.
Future work should further analyze the perspectives of model developers and model users to design effective solutions that make ethical considerations more tangible in practice.

\section*{Acknowledgment}

We acknowledge DASHH, Data Science in Hamburg -- Helmholtz Graduate School for the Structure of Matter, for financial support.

%% file: tables/data.tex
\begin{table} 
    \centering
    \caption{analyzed artifacts divided into ai guidelines, ai~documentation frameworks, and model cards studies.}
    \rowcolors{2}{white}{gray!20} 
    \renewcommand{\arraystretch}{1.2}
    \begin{tabular}{>{\raggedright\arraybackslash}p{0.21\columnwidth} >{\raggedright\arraybackslash}p{0.69\columnwidth}}
         \hline\noalign{\vskip 2pt}
         Data Source & Analyzed Artifacts \\
         \noalign{\vskip 2pt}\hline
         ACM Digital Library \textbf{(A)} &  
         Iliadis~\cite{iliadis_ai-gfa_2024},
         Giudici et al.~\cite{giudici_artificial_2024},
         Khalid et al.~\cite{khalid_privacy-preserving_2023},
         Liu et al.~\cite{liu_enhancing_2024},
         Diáz-Rodríguez et al.~\cite{diaz-rodriguez_connecting_2023}\\
         IEEE Xplore \textbf{(B)} & 
         Golbin et al.~\cite{golbin_responsible_2020},
         Lu et al.~\cite{lu_responsible-ai-by-design_2023},
         Lu et al.~\cite{lu_towards_2022},
         Sanderson et al.~\cite{sanderson_towards_2022},
         Zhang et al.~\cite{zhang_trusted_2021} \\
         SpringerLink \textbf{(C)} & 
         Akula and Garibay~\cite{akula_ethical_2021},
         Leopold~\cite{leopold_mastering_2023},
         Slavkovik~\cite{slavkovik_mythical_2023},
         Boddington~\cite{boddington_rise_2023},
         Akhtar et al.~\cite{akhtar_transparency_2024}\\
         Elsevier Scopus \textbf{(D)}& 
         Kuwajima and Ishikawa~\cite{kuwajima_adapting_2019},
         Jedličková~\cite{jedlickova_ethical_2024},
         López et al.~\cite{lopez_requirements_2024},
         Hartikainen et al.~\cite{hartikainen_towards_2023},
         Lier et al.~\cite{lier_what_2024}\\
         Google Scholar \textbf{(E)} & 
         Ryan and Stahl~\cite{ryan_artificial_2021},
         Keskinbora~\cite{keskinbora_medical_2019},
         Ghallab~\cite{ghallab_responsible_2019},
         Hagendorff~\cite{hagendorff_ethics_2020},
         Jobin et al.~\cite{jobin_global_2019}\\
         European Com- mission \textbf{(F)} & 
         European Commission~\cite{eu_2020}\\
         \hline
         \gls{ai} documenta- tion frameworks \textbf{(G-I)} & 
         Mitchell et al.~\cite{mitchell_model_2019},
         Arnold et al.~\cite{arnold_factsheets_2019},
         Sokol and Flach~\cite{sokol_explainability_2020}\\
         \hline
         Model card analysis papers \textbf{(J)} & 
         Gao et al.~\cite{gao_documenting_2024},
         Pepe et al.~\cite{pepe_how_2024},
         Bhat et al.~\cite{bhat_aspirations_2023}\\
         Model cards (recommended) \textbf{(K)} & 
         databricks/dolly-v2-7b~\cite{databricks_databricksdolly-v2-7b_nodate},
         openai/whisper-medium~\cite{openai_openaiwhisper-medium_nodate},
         optimum/distilbert-base-uncased-finetuned-sst-2-english~\cite{hugging_face_optimum_optimumdistilbert-base-uncased-finetuned-sst-2-english_nodate},
         Salesforce/ctrl~\cite{salesforce_salesforcectrl_nodate},
         tiiuae/falcon-180B~\cite{technology_innovation_institute_tiiuaefalcon-180b_nodate}\\
         Model cards (trending) \textbf{(L)} & 
         agentica-org/DeepScaleR-1.5B-Preview~\cite{agentica_agentica-orgdeepscaler-15b-preview_nodate},
         black-forest-labs/FLUX.1-dev~\cite{black_forest_labs_black-forest-labsflux1-dev_nodate},
         nomic-ai/nomic-embed-text-v2-moe~\cite{nomic_ai_nomic-ainomic-embed-text-v2-moe_nodate},
         NousResearch/DeepHermes-3-Llama-3-8B-Preview~\cite{nousresearch_nousresearchdeephermes-3-llama-3-8b-preview_nodate},
         perplexity-ai/r1-1776~\cite{perplexity_perplexity-air1-1776_nodate}\\
         \noalign{\vskip 2pt}\hline
    \end{tabular}
    \label{tab:data}
\end{table}

%% file: tables/taxonomy.tex
\begin{table*}
\centering
\caption{taxonomy of ethical requirements for ai models based on the analysis of ai guidelines (a-f), ai documentation frameworks (g-i), and model cards (j-l). a-e and j-l are displayed as bins per data source (see table \ref{tab:data}); a full version is available in the replication package \cite{rep_pack}. matrix values are totals of binary values (1 or 0) representing the presence/absence of a requirement per source.}
\rowcolors{3}{white}{gray!20} 
\renewcommand{\arraystretch}{1.1}
\begin{tabular}{p{10.8cm}|p{0.1cm}p{0.1cm}p{0.1cm}p{0.1cm}p{0.1cm}p{0.2cm}|p{0.1cm}p{0.1cm}p{0.15cm}|p{0.1cm}p{0.1cm}p{0.3cm}}
\hline
& \multicolumn{6}{c|}{Guidelines} & \multicolumn{3}{c|}{Framew.} & \multicolumn{3}{c}{Model Ca.} \\
\textbf{Ethical Principle}, \textbf{\textit{Sub-Principle}}, Requirement & A & B & C & D & E & F & G & H & I & J & K & L \\
\hline
\multicolumn{13}{l}{\textbf{Reliability}} \\
\multicolumn{13}{l}{\textbf{\textit{Accuracy: How do developers ensure continuously correct model output?}}} \\
(R01) Explain quality of training/evaluation data, process, and results & 1 & 3 & 2 & 2 & 2 & 1 & 1 & 1 & 1 & 2 & 5 & 5 \\
(R02) Explain mechanisms in place to continuously ensure accuracy & 1 & 4 &  & 1 & 3 & 1 &  & 1 &  &  &  &  \\
(R03) Explain factors that influence model accuracy &  &  & 1 & 1 &  &  & 1 &  &  & 1 &  &  \\
\multicolumn{13}{l}{\textbf{\textit{Safety and Security: How do developers ensure protection of individuals and groups from harm?}}} \\
(R04) Explain security measures for mitigating external threats & 3 & 4 & 2 & 2 & 4 & 1 & 1 & 1 & 1 &  &  &  \\
(R05) Explain safety measures for increasing robustness and mitigating harm & 3 & 4 & 3 & 3 & 3 & 1 & 1 & 1 &  & 1 & 1 &  \\
(R06) Explain measures for protecting privacy of data subjects & 2 & 5 & 2 & 3 & 4 & 1 & 1 & 1 &  &  &  &  \\
(R07) Explain duration of safety and security coverage &  &  &  &  &  & 1 &  &  &  &  &  &  \\
(R08) Provide risk assessment for model usage & 3 & 2 &  & 2 & 2 & 1 & 1 & 1 &  & 3 & 4 & 2 \\
(R09) Provide recommendations for users to mitigate risks &  &  &  &  &  &  & 1 &  &  & 1 & 4 &  \\
\multicolumn{13}{l}{\textbf{\textit{Auditability: How do developers ensure that the model qualities are reviewed and verified?}}} \\
(R10) Explain measures in place to ensure auditability & 1 &  &  & 3 &  & 1 &  &  &  &  &  &  \\
(R11) Provide details on audit procedures and results & 1 & 1 & 1 & 2 & 2 & 1 & 1 & 1 &  & 1 & 1 &  \\
(R12) Provide certification to demonstrate compliance with laws, policies, and standards &  & 1 & 1 &  & 1 & 1 &  &  &  &  &  &  \\
\hline
\multicolumn{13}{l}{\textbf{Transparency}} \\
\multicolumn{13}{l}{\textbf{\textit{Communication of Capabilities: How do developers ensure to communicate what the model can and cannot do?}}} \\
(T01) Explain intended uses and applicable domains & 1 & 1 &  & 2 & 1 & 1 & 1 & 1 & 1 & 1 & 4 & 5 \\
(T02) Explain out-of-scope uses and model limitations &  & 1 &  & 1 & 1 & 1 & 1 & 1 &  & 2 & 4 & 2 \\
(T03) Provide details on model architecture and algorithms &  & 1 & 1 & 1 & 1 &  & 1 & 1 & 1 & 1 & 4 & 3 \\
(T04) Provide training and evaluation data &  & 1 &  & 1 &  &  & 1 & 1 &  & 2 & 3 & 2 \\
(T05) Explain trade-offs made for competing design goals & 3 &  & 1 & 1 &  & 1 &  & 1 &  &  &  &  \\
(T06) Explain how users should correctly interact with the model &  &  & 1 &  &  & 1 &  & 1 &  & 1 & 4 & 3 \\
(T07) Refer to additional documentation &  &  &  &  &  &  & 1 &  &  & 1 & 5 & 5 \\
\multicolumn{13}{l}{\textbf{\textit{Explainability: How do developers ensure interpretable model predictions?}}} \\
(T08) Explain how and for whom the model generates explanations & 4 & 4 & 5 & 3 & 4 & 1 &  & 1 & 1 &  &  &  \\
(T09) Explain how explanation quality was tested &  &  &  &  &  &  &  & 1 & 1 &  &  &  \\
\multicolumn{13}{l}{\textbf{\textit{Traceability: How do developers ensure documented and reproducible model development and predictions?}}} \\
(T10) Explain where, how, and when training data was obtained and processed &  &  & 1 & 1 & 1 & 1 &  &  &  &  &  &  \\
(T11) Explain how model and data modifications are logged &  & 4 & 1 & 1 &  & 1 & 1 & 1 &  & 1 &  &  \\
(T12) Explain how user interaction data is logged & 1 & 3 &  & 1 & 1 & 1 &  &  &  &  &  &  \\
\hline
\multicolumn{13}{l}{\textbf{Empowerment}} \\
\multicolumn{13}{l}{\textbf{\textit{User Autonomy: How do developers ensure empowering rather than restricting or manipulating users?}}} \\
(E01) Explain how the user is informed about interacting with an AI & 1 & 1 &  & 2 & 2 & 1 &  &  &  &  &  &  \\
(E02) Explain how the model assists users in decision-making without manipulation & 1 & 1 &  & 3 & 2 & 1 &  &  &  &  &  &  \\
(E03) Explain how human oversight can influence or override model predictions & 2 & 1 & 1 & 3 &  & 1 &  &  &  &  &  &  \\
\multicolumn{13}{l}{\textbf{\textit{Consent and Control: How do developers ensure the authority of data subjects?}}} \\
(E04) Explain how data subjects are informed of data processing and can manage consent & 1 &  & 2 & 2 & 2 & 1 &  & 1 &  & 1 &  &  \\
(E05) Explain how data subjects can access and request the change or deletion of their data & 1 &  &  & 1 & 1 & 1 &  & 1 &  &  &  &  \\
\multicolumn{13}{l}{\textbf{\textit{Liability: How do developers ensure their accountability for the model?}}} \\
(E06) Name responsible and liable entities throughout the model life cycle & 1 & 3 & 1 &  & 2 & 1 & 1 & 1 &  & 1 &  & 1 \\
(E07) Explain procedures for contesting and redressing harmful model behavior & 1 & 2 & 2 &  & 2 & 1 &  & 1 &  & 1 &  &  \\
(E08) Explain procedures for handling liability claims &  &  &  &  & 1 &  &  & 1 &  &  &  &  \\
(E09) Explain how model developers report harm to affected parties and the public &  &  &  &  & 1 & 1 &  &  &  &  &  &  \\
(E10) Explain which responsibilities the model developers disclaim &  &  &  &  &  &  & 1 &  &  & 3 &  &  \\
\hline
\multicolumn{13}{l}{\textbf{Beneficence}} \\
\multicolumn{13}{l}{\textbf{\textit{Fairness: How do developers ensure non-discriminatory, equitable model output?}}} \\
(B01) Explain how diverse stakeholders were involved in model development & 3 & 2 & 1 & 3 & 2 & 1 & 1 & 1 &  &  &  &  \\
(B02) Explain bias detection techniques applied to datasets (pre-processing) & 2 & 3 & 2 & 3 & 4 & 1 & 1 & 1 &  & 1 &  &  \\
(B03) Explain bias detection techniques applied during model training (in-processing) & 2 & 2 & 1 & 1 & 1 & 1 &  &  &  &  &  &  \\
(B04) Explain bias detection techniques applied to model output (post-processing) & 1 & 1 & 2 & 2 &  & 1 & 1 & 1 &  &  &  &  \\
\multicolumn{13}{l}{\textbf{\textit{Social Beneficence: How do developers ensure positive impact of the model on the well-being of society?}}} \\
(B05) Explain positive impact the model can have on individuals and society & 1 &  & 2 & 4 & 2 & 1 &  &  &  & 1 & 2 &  \\
(B06) Explain usage permissions and restrictions &  &  &  &  & 1 &  & 1 &  &  & 3 & 3 & 2 \\
\multicolumn{13}{l}{\textbf{\textit{Environmental Beneficence: How do developers ensure environmental sustainability?}}} \\
(B07) Explain positive impact the model can have on the environment & 1 &  & 1 &  & 2 & 1 &  &  &  &  &  &  \\
(B08) Provide measured values for environmental impact metrics & 1 &  &  &  &  & 1 &  &  &  & 1 &  &  \\
(B09) Explain efforts made to minimize the environmental footprint of the model & 2 &  &  &  & 2 & 1 &  &  &  &  &  & 1 \\
\hline
\textit{Number of distinct requirements covered per data source} & 27 & 24 & 23 & 28 & 29 & 34 & 20 & 24 & 6 & 21 & 13 & 11\\
\hline
\end{tabular}
\label{tab:taxonomy}
\end{table*}